\documentclass{ws-procs9x6-cpt16}
\begin{document}

\def\etal {{\it et al.}}

\title{Recent Progress in Lorentz and CPT Violation}

\author{V.\ Alan Kosteleck\'y}

\address{Physics Department, Indiana University\\
Bloomington, IN 47405, USA}

\begin{abstract}
This contribution to 
the CPT'16 meeting
briefly highlights some of the recent progress 
in the phenomenology of Lorentz and CPT violation,
with emphasis on research 
performed at the Indiana University Center for Spacetime Symmetries.
\end{abstract}

\bodymatter

\section{Introduction}

In the three years since the CPT'13 meeting,
interest in the idea of Lorentz and CPT violation has continued unabated,
driven by the notion that tiny detectable violations of Lorentz symmetry 
could yield experimental information about Planck-scale physics.
New results drawn from many subfields
are appearing on the timescale of weeks,
often announcing sensitivity gains of an order of magnitude or more,
and making the subject 
among the more rapidly developing areas of physics.
One simple measure of the rate of progress during these three years 
is the increase of over 40\% in length of the
{\it Data Tables for Lorentz and CPT Violation},\cite{tables}
which collates experimental measurements in all subfields.
Here,
I outline a few essentials of this active subject
and briefly highlight some of the recent research performed
at the Indiana University Center for Spacetime Symmetries (IUCSS).

\section{Essentials}

At least two philosophically different approaches can be envisaged
in describing a new phenomenon.
One is to develop a specific model and study its predictions.
This method is well suited to situations
where an experimental effect is confirmed.
However,
given the present lack of compelling evidence for Lorentz violation,
it is appropriate to adopt a more general alternative method,
developing a realistic framework encompassing
all violations of Lorentz and CPT symmetry
to guide a broad experimental search.

Effective field theory provides
a potent tool for describing small signals emerging
from an otherwise inaccessible scale.\cite{eft}
The comprehensive effective field theory for Lorentz violation
that integrates coordinate independence, realism, and generality
is called the Standard-Model Extension (SME).\cite{ck,akgrav}
The SME can be constructed
from the action of General Relativity coupled to the Standard Model
by adding all Lorentz-violating and coordinate-independent terms.
These can arise explicitly or spontaneously 
in a unified theory such as strings,\cite{ksp}
and they incorporate general CPT violation.\cite{ck,owg}

Each SME term comes with a coefficient for Lorentz violation  
governing the size of the associated experimental signals.
A coefficient can be viewed as a background
that affects the properties of particles 
according to their flavor, velocity, spin, and couplings.
The effects of the coefficients are expected to be tiny,
either through direct suppression
or through `countershading' by naturally weak couplings.\cite{akjt}
Experimental constraints now exist for many coefficients,\cite{tables}
some at Planck-suppressed levels or beyond.

Terms in the SME Lagrange density include
Lorentz-violating operators of arbitrary mass dimension $d$.
Restricting attention to operators of renormalizable dimension $d\leq 4$
yields the so-called minimal SME.\cite{ck,akgrav}
The explicit construction of the numerous operators for arbitrary $d>4$ 
has been accomplished for several sectors,
including terms associated with the propagation of 
gravitons,
photons,
Dirac fermions,
and neutrinos.\cite{km}

To preserve conventional Riemann or Riemann-Cartan geometry,
the Lorentz violation must be spontaneous.\cite{akgrav} 
This implies that massless Nambu-Goldstone modes appear,
with accompanying phenomenological effects.\cite{lvng}
The conjecture\cite{akgrav} that explicit Lorentz violation
is associated instead with Finsler geometry
has gained recent support\cite{finsler}
but remains open to date.

\section{Recent IUCSS progress}

At the IUCSS,
much of the focus during the last three years
has been on the gravity, photon, matter, and quark sectors.
In the gravity sector,
the nonminimal pure-gravity terms for $d\leq 6$
have recently been constructed.\cite{bkx}
They modify gravity at short distances,
and in the nonrelativistic limit the effects are controlled
by 14 independent coefficients. 
In a series of experimental advances during 2015 and 2016,
the first combined sensitivities to these coefficients were reported,
and individual bounds then further improved 
by about two orders of magnitude.\cite{shortrange}
In a different vein, 
all contributions to the graviton propagator were constructed
for arbitrary $d$,\cite{km}
including both Lorentz-invariant and Lorentz-violating terms.
These reveal anisotropic, dispersive, and birefringent modifications
to gravitational-wave propagation,
which are constrained partly by observational data.
Another source of strong bounds on SME coefficients
in the pure-gravity sector
comes from the highest-energy cosmic rays,
which constrain gravitational \v{C}erenkov radiation.\cite{kt15}
Nonetheless,
much of the coefficient space in the gravity sector
remains open for future exploration.

Tests of Lorentz symmetry with light and matter have
the longest history but continue to set record sensitivities.
In the minimal photon sector,
an improvement in sensitivity of some four orders of magnitude
has been achieved in recent months
using data from one of the LIGO instruments,\cite{kmm16}
showing that Planck-suppressed effects on the propagation of light
can be accessed by the world's largest interferometers.
In the nonminimal photon sector,
astrophysical measurements have bounded 
individually all $d=6$ nonbirefringent effects,\cite{ki15}
and results constraining all $d=5$ coefficients 
are within reach.
The phenomenology of the nonminimal matter sector 
has also recently seen significant progress.
Spectroscopic methods for hydrogen, antihydrogen,
other hydrogenic systems, and exotic atoms
provide constraints from existing data
and offer access to many unmeasured coefficients,
as do studies of particles confined to a Penning trap.\cite{amo}

In the quark sector,
limits on Lorentz and CPT violation 
are comparatively few to date.
Most have been obtained from meson interferometry,
which offers a unique sensitivity to certain quark coefficients.\cite{ak98}
The past few years have seen 
improved measurements on $d$- and $s$-quark coefficients using kaons
and first bounds on $b$ quarks 
from both $B_d$ and $B_s$ mixing.\cite{mesonexpt} 
The $t$ quark decays too rapidly for mixing
and hence its Lorentz properties were unknown until recently,
when studies of $t$-$\overline{t}$ pair production and decay
yielded first constraints on $t$-quark coefficients.\cite{d0top} 
It has now been shown that this result could be improved 
in experiments at the Large Hadron Collider,
and the first test of CPT in the top sector could be performed
using single-$t$ production.\cite{bkl16}
Different avenues to investigating the quark sector
are also being explored.
One is using deep inelastic scattering,
from which bounds on $u$- and $d$-quark coefficients
can be extracted.\cite{klv16} 
Another is using chiral perturbation theory,
which can connect hadron coefficients to quark coefficients.\cite{lvchpt}
Both these approaches offer the potential 
for a significant expansion of our understanding
of Lorentz and CPT violation in quarks. 
The prospects for future discovery are bright.

\section*{Acknowledgments}

This work was supported in part
by the U.S.\ D.o.E.\ grant {DE}-SC0010120
and by the Indiana University Center for Spacetime Symmetries.

\end{document}